\documentclass[11pt]{article}
\usepackage{graphicx,bm}

\def\k{{\bm k}}
\def\u{{\bm u}}
\def\x{{\bm x}}
\def\bomega{{\bm \omega}}

\def\norm#1{\left|\mkern-2mu\left|#1\right|\mkern-2mu\right|}
\def\norm#1{\left|\mkern-2mu\left|#1\right|\mkern-2mu\right|}

\begin{document}

\begin{center}{{\large \bf Constraints on scalar diffusion anomaly in 
three-dimensional flows having bounded velocity gradients}\\~\\
Chuong V. Tran\footnote{chuong@mcs.st-and.ac.uk}\\
School of Mathematics and Statistics, University of St Andrews\\
St Andrews KY16 9SS, United Kingdom}
\end{center}
\date{\today}

\baselineskip=16pt

\centerline{\bf Abstract}

This study is concerned with the decay behaviour of a passive scalar 
$\theta$ in three-dimensional flows having bounded velocity gradients. 
Given an initially smooth scalar distribution, the decay rate 
$d\langle\theta^2\rangle/dt$ of the scalar variance 
$\langle\theta^2\rangle$ is found to be bounded in terms of controlled 
physical parameters. Furthermore, in the zero diffusivity limit, 
$\kappa\to0$, this rate vanishes as $\kappa^{\alpha_0}$ if there 
exists an $\alpha_0\in(0,1]$ independent of $\kappa$ such that 
$\langle|(-\Delta)^{\alpha/2}\theta|^2\rangle<\infty$ for 
$\alpha\le\alpha_0$. This condition is satisfied if in the limit 
$\kappa\to0$, the variance spectrum $\Theta(k)$ remains steeper than 
$k^{-1}$ for large wave numbers $k$. When no such positive $\alpha_0$ 
exists, the scalar field may be said to become virtually singular. A 
plausible scenario consistent with Batchelor's theory is that $\Theta(k)$ 
becomes increasingly shallower for smaller $\kappa$, approaching the 
Batchelor scaling $k^{-1}$ in the limit $\kappa\to0$. For this classical 
case, the decay rate also vanishes, albeit more slowly --- like 
$(\ln P_r)^{-1}$, 
where $P_r$ is the Prandtl or Schmidt number. Hence, diffusion anomaly 
is ruled out for a broad range of scalar distribution, including 
power-law spectra no shallower than $k^{-1}$. The implication is that 
in order to have a $\kappa$-independent and non-vanishing decay rate, 
the variance at small scales must necessarily be greater than that 
allowed by the Batchelor spectrum. These results are discussed in the 
light of existing literature on the asymptotic exponential decay 
$\langle\theta^2\rangle\sim e^{-\gamma t}$, where $\gamma>0$ is
independent of $\kappa$. 

~

\centerline{xxxxxxxxxxxxxxxxxxxxxxxxx}

\section{Introduction}

The transport and diffusion of a passive scalar $\theta(\x,t)$ in  
incompressible fluid flows $\u(\x,t)$ is governed by the 
advection-diffusion equation,
\begin{eqnarray}
\label{governing}
\partial_t\theta + \u\cdot\nabla\theta &=& \kappa\Delta\theta,
~~~~\nabla\cdot\u=0,
\end{eqnarray}
where $\kappa$ is the diffusivity. The incompressible flows $\u$ may be 
described by a separate system of equations, for example the Navier--Stokes 
system, or just satisfy some prescribed conditions. This study considers 
three-dimensional flows having bounded velocity gradients, i.e.
$|\nabla\u|<\infty$. For simplicity, periodic boundary conditions are 
imposed on both $\u$ and $\theta$, and the scalar spatial average is set 
to zero, i.e. $\langle\theta\rangle=0$. The initial scalar distribution 
is assumed to be smooth. The diffusive decay of the scalar variance 
$\langle\theta^2\rangle$ is governed by
\begin{eqnarray}
\label{decay}
\frac{d}{dt}\langle\theta^2\rangle &=& 
-2\kappa\langle|\nabla\theta|^2\rangle.
\end{eqnarray}
A problem of long-standing interest is to what extent this decay can be 
accelerated as the scalar field becomes ``less smooth'' through advective 
amplification of its gradients. 

This problem, together with related issues concerning the scalar spectrum, 
has been actively studied for decades,$^{1-21}$ most notably since the 
seminal work of Batchelor and coworkers.$^{3,4}$ In one of these studies, 
Batchelor$^3$ considered presumably non-singular Navier--Stokes flows, 
where the fluid viscosity $\nu$ is much greater than the scalar diffusivity 
$\kappa$. This so-called regime of large Prandtl or Schmidt number $P_r$, 
defined by $P_r=\nu/\kappa$, is of both practical and theoretical interest, 
where a fundamental question is whether the decay rate 
$d\langle\theta^2\rangle/dt$ would remain non-zero in the limit 
$\kappa\to0$. The general belief, having its root in the phenomenological 
theory of turbulence, is that this rate becomes independent of $\kappa$ 
and remains non-zero in that limit (even for two-dimensional flows, which 
are presumably poorer stirrers as compared with their three-dimensional
counterparts). Thus, the scalar variance is believed to cascade to the small 
scales and undergo anomalous diffusion just as the fluid kinetic energy. 
However, doubts have been cast upon this picture.$^{11,16}$ In particular, 
Sreenivasan$^{16}$ suggested that the analogy between the energy and 
scalar variance is rather weak and questioned whether there is a passive 
scalar cascade at all (for a discussion see Chapter 5 of Davidson's 
book).$^{22}$ Recently, Tran and Dritschel$^{23}$ showed that for general 
two-dimensional flows having finite enstrophy and power-law spectra of 
$\langle\theta^2\rangle$ no shallower than $k^{-1}$, the decay rate 
$2\kappa\langle|\nabla\theta|^2\rangle$ vanishes in the limit 
$\kappa\to0$. Their calculation took full account of the suppression 
of scalar gradients by diffusive effects. 

This study extends the above result to the three-dimensional case, with
an improvement in the method of analysis by making use of an analytic 
inequality. For flows having bounded velocity gradients, the maximally 
achievable decay rate $2\kappa\langle|\nabla\theta|^2\rangle$ is shown 
to be bounded from above. Furthermore, in the limit $\kappa\to0$, the 
derived upper bound vanishes provided that the scalar field remains 
marginally smooth. More precisely, if there exists a positive $\alpha_0$ 
independent of $\kappa$ such that 
$\langle|(-\Delta)^{\alpha/2}\theta|^2\rangle<\infty$ for 
$\alpha\le\alpha_0$, then 
$\kappa\langle|\nabla\theta|^2\rangle\le C\kappa^{\alpha_0}$.
Here, $\alpha_0\in(0,1]$ and $C$ is a constant, depending on 
$|\nabla\u|$ and $\langle|(-\Delta)^{\alpha_0/2}\theta|^2\rangle$, 
both being bounded and independent of $\kappa$. This constraint covers 
all scalar distribution having variance spectrum $\Theta(k)$ steeper than 
$k^{-1}$ at the spectral tail, i.e. at large wave numbers $k$. This is 
an improvement over the result of Tran and Dritschel$^{23}$ for the 
two-dimensional case, where power-law spectra were assumed explicitly in 
their calculations. When no such positive $\alpha_0$ exists, a plausible 
scenario consistent with Batchelor's theory$^{3}$ is 
that $\Theta(k)$ becomes increasingly shallower for smaller $\kappa$, 
approaching the Batchelor scaling $k^{-1}$ in the limit $\kappa\to0$. For 
this classical case, the semi-analytic estimate of Tran and Dritschel can 
be applied, yielding a vanishing decay rate no slower than $(\ln P_r)^{-1}$. 
Hence, diffusion anomaly is ruled out for a broad class of scalar 
distribution, including power-law spectra no shallower than $k^{-1}$. The 
implication is that a non-vanishing decay rate in the limit $\kappa\to0$ 
necessarily requires the variance at small scales to be greater than that 
allowed by the Batchelor spectrum. This is consistent with recent 
studies$^{1,7,14,19}$ suggesting spectra shallower than $k^{-1}$. This
analysis when applied to the forced case implies that for power-law 
$\Theta(k)$, the scalar variance diverges in the limit $\kappa\to0$.
Finally, the present results are discussed in the light of existing
literature$^{8,13,17,20}$ on the exponential decay behavior at large times,  
when the exponential decay rate becomes independent of $\kappa$ and 
remains non-zero. 

\section{Basic estimates}

This section briefly recalls a basic calculus inequality and related 
estimates, both being used in this study. In what follows, $\theta$ 
is bounded and assumed to be sufficiently smooth so that the quantities 
under consideration are well defined. 

The Fourier representation of $\theta(\x,t)$ is
\begin{eqnarray}  
\label{fourier}
\theta(\x) = \sum_{\k}\widehat{\theta}(\k)\exp\{i\k\cdot\x\},
\end{eqnarray}
where $\widehat{\theta}(\k)$ is the Fourier transform of $\theta(\x)$
and $\k$ is the wave vector. In (\ref{fourier}), the time variable has 
been suppressed for convenience. For real $\alpha$, including negative 
values, the fractional derivative $(-\Delta)^{\alpha/2}$ is defined by
\begin{eqnarray}  
\label{fractional}
(-\Delta)^{\alpha/2}\theta(\x) = 
\sum_{\k}k^\alpha\widehat{\theta}(\k)\exp\{i\k\cdot\x\},
\end{eqnarray} 
where $k=|\k|$ is the wave number. For $\alpha\le1$, the following 
interpolation-type inequality holds
\begin{eqnarray}
\label{holder}
\langle|\nabla\theta|^2\rangle 
&\le&
\langle|\Delta\theta|^2\rangle^{(1-\alpha)/(2-\alpha)} 
\langle|(-\Delta)^{\alpha/2}\theta|^2\rangle^{1/(2-\alpha)}.
\end{eqnarray}
This inequality reduces to the Cauchy--Schwarz inequality and an 
identity for $\alpha=0$ and $\alpha=1$, respectively. For $\alpha<1$, 
Eq. (\ref{holder}) can be proved by an elementary method,$^{24,25}$ 
and the proof goes as follows. One has
\begin{eqnarray}
\label{holder1}
\int k^2\Theta(k)\,dk &=& 
\int(k^4\Theta(k))^{(1-\alpha)/(2-\alpha)}
(k^{2\alpha}\Theta(k))^{1/(2-\alpha)}\,dk \nonumber\\
&\le&
\left(\int k^4\Theta(k)\,dk\right)^{(1-\alpha)/(2-\alpha)}
\left(\int k^{2\alpha}\Theta(k)\,dk\right)^{1/(2-\alpha)},
\end{eqnarray}
where the H\"older inequality with the pair of conjugate exponents
$p=(2-\alpha)/(1-\alpha)$ and $q=2-\alpha$ has been used. Eq. (\ref{holder1}) 
is equivalent to (\ref{holder}), thus proving (\ref{holder}). For 
convenient application in the subsequent sections, let us rewrite 
(\ref{holder}) as $\langle|\nabla\theta|^2\rangle^{(2-\alpha)/(1-\alpha)} 
\le\langle|\Delta\theta|^2\rangle
\langle|(-\Delta)^{\alpha/2}\theta|^2\rangle^{1/(1-\alpha)}$ 
and re-arrange the factors to obtain
\begin{eqnarray}
\label{holder2}
\frac{\langle|\nabla\theta|^2\rangle^2}{\langle|\Delta\theta|^2\rangle}
&\le& 
\frac{\langle|(-\Delta)^{\alpha/2}\theta|^2\rangle^{1/(1-\alpha)}}
{\langle|\nabla\theta|^2\rangle^{\alpha/(1-\alpha)}}.
\end{eqnarray}
Equation (\ref{holder2}) implies that if there exists an $\alpha_0>0$ 
such that $\langle|(-\Delta)^{\alpha_0/2}\theta|^2\rangle<\infty$ for 
$\alpha\le\alpha_0$, then the ratio on its left-hand side vanishes in 
the limit $\langle|\nabla\theta|^2\rangle\to\infty$ because the right-hand 
side vanishes in this limit. This is the case if $\Theta(k)$ is steeper 
than $k^{-1}$ for large $k$.

For the critical spectrum $\Theta(k)\propto k^{-1}$ extending to infinity,
no such $\alpha_0$ exists. Nevertheless, the ratio 
$\langle|\nabla\theta|^2\rangle^2/\langle|\Delta\theta|^2\rangle$
also vanishes for bounded $\langle\theta^2\rangle$. Indeed, if 
$\Theta(k)=ck^{-1}$ for $k\in[k_1,k_2]$ (where $k_1\ll k_2$) and 
negligible otherwise (steeper than $k^{-1}$ for $k>k_2$), then 
$c=\langle\theta^2\rangle/\ln(k_2/k_1)$. It follows that
\begin{eqnarray}
\label{holderer}
\frac{\langle|\nabla\theta|^2\rangle^2}{\langle|\Delta\theta|^2\rangle}
&=& 
\frac{\left(c\int_{k_1}^{k_2}k\,dk\right)^2}
{c\int_{k_1}^{k_2}k^3\,dk}
=\frac{\langle\theta^2\rangle}{\ln(k_2/k_1)}.
\end{eqnarray}
The limit $k_2\to\infty$ necessarily entails $\ln(k_2/k_1)\to\infty$ 
if $k_1$ remains bounded. Hence 
$\langle|\nabla\theta|^2\rangle^2/\langle|\Delta\theta|^2\rangle\to0$
as claimed. Note that the inclusion of the contribution from $k>k_2$, 
where $\Theta(k)$ is steeper than $k^{-1}$, does not alter (\ref{holderer}) 
in any significant manner (see Tran and Dritschel$^{23}$ for details). 

The above results imply that given a finite field $\theta$, the ratio 
$\langle|\nabla\theta|^2\rangle^2/\langle|\Delta\theta|^2\rangle$
vanishes for a broad range of spectral distribution having divergent
mean-square gradients, $\langle|\nabla\theta|^2\rangle\to\infty$. This 
includes all $\Theta(k)$ no shallower than $k^{-1}$ for large $k$. This 
type of distribution may be called ``regular''. The 
type of ``irregular'' distribution that gives rise to a non-zero ratio 
$\langle|\nabla\theta|^2\rangle^2/\langle|\Delta\theta|^2\rangle$ 
in the limit $\langle|\nabla\theta|^2\rangle\to\infty$ includes power-law 
spectra shallower than $k^{-1}$ and non-power-law spectra having a 
significant fraction of $\langle\theta^2\rangle$ at small scales. This 
fraction must be greater than that allowed by a $k^{-1}$ spectrum and 
non-diminishing. A more quantitative sense of this fraction is given in 
what follows. For a finite (but otherwise arbitrarily large) wave number 
$k_*$, let $\theta=\theta_<+\theta_>$, where the large-scale component 
$\theta_<$ and small-scale component $\theta_>$ are defined by
\begin{eqnarray} 
\theta_<=\sum_{k<k_*}\widehat{\theta}(\k)\exp\{i\k\cdot\x\}
\end{eqnarray}
and
\begin{eqnarray} 
\theta_> = \sum_{k\ge k_*}\widehat{\theta}(\k)\exp\{i\k\cdot\x\},
\end{eqnarray}
respectively. In terms of these components, the ratio 
$\langle|\nabla\theta|^2\rangle^2/\langle|\Delta\theta|^2\rangle$
is given by
\begin{eqnarray}
\label{singular1}
\frac{\langle|\nabla\theta|^2\rangle^2}{\langle|\Delta\theta|^2\rangle}
&=&
\frac{\left(\langle|\nabla\theta_<|^2\rangle +
\langle|\nabla\theta_>|^2\rangle\right)^2}
{\langle|\Delta\theta_<|^2\rangle+\langle|\Delta\theta_>|^2\rangle} \nonumber\\
&=&
\frac{\left(\frac{\langle|\nabla\theta_<|^2\rangle}
{\langle|\Delta\theta_>|^2\rangle^{1/2}}+
\frac{\langle|\nabla\theta_>|^2\rangle}
{\langle|\Delta\theta_>|^2\rangle^{1/2}}\right)^2}
{\frac{\langle|\Delta\theta_<|^2\rangle}
{\langle|\Delta\theta_>|^2\rangle}+1}.
\end{eqnarray}
Now, since $\langle|\nabla\theta_<|^2\rangle\le k_*^2\langle\theta_<^2\rangle
<\infty$, the limit $\langle|\nabla\theta|^2\rangle\to\infty$ requires 
$\langle|\nabla\theta_>|^2\rangle\to\infty$, which in turn entails
$\langle|\Delta\theta_>|^2\rangle\to\infty$ because 
$\langle|\Delta\theta_>|^2\rangle\ge k_*^2\langle|\nabla\theta_>|^2\rangle$.
Furthermore, 
$\langle|\Delta\theta_<|^2\rangle\le k_*^4\langle\theta_<^2\rangle<\infty$.
It follows that both 
$\langle|\nabla\theta_<|^2\rangle/\langle|\Delta\theta_>|^2\rangle^{1/2}\to0$
and $\langle|\Delta\theta_<|^2\rangle/\langle|\Delta\theta_>|^2\rangle\to0$.
Hence, if the ratio
$\langle|\nabla\theta|^2\rangle^2/\langle|\Delta\theta|^2\rangle$ 
remains non-zero as $\langle|\nabla\theta|^2\rangle\to\infty$, say 
bounded away from zero by $\epsilon>0$, then from (\ref{singular1}), 
it is readily deduced that
\begin{eqnarray}
\label{singular2}
\epsilon \le \frac{\langle|\nabla\theta|^2\rangle^2}
{\langle|\Delta\theta|^2\rangle} \longrightarrow
\frac{\langle|\nabla\theta_>|^2\rangle^2}
{\langle|\Delta\theta_>|^2\rangle}.
\end{eqnarray}
It follows that
\begin{eqnarray}
\label{singular3}
\langle\theta_>^2\rangle \ge
\frac{\langle|\nabla\theta_>|^2\rangle^2}
{\langle|\Delta\theta_>|^2\rangle} \ge \epsilon,
\end{eqnarray}
where the left-hand side inequality is the familiar Cauchy-schwarz inequality
for the small-scale component $\theta_>$. Eq. (\ref{singular3}) gives 
an explicit bound for $\langle\theta_>^2\rangle$ in terms of $\epsilon$. 
Thus, the variance at small scales $k\ge k_*$ is bounded away from zero 
if the ratio 
$\langle|\nabla\theta|^2\rangle^2/\langle|\Delta\theta|^2\rangle$ is
to remain non-zero in the limit $\langle|\nabla\theta|^2\rangle\to\infty$.

\section{Main results}

This section derives the main results of this study, including the proof 
of absence of diffusion anomaly for a broad range of scalar distribution 
mentioned earlier in \S\,1. The results are then discussed in the light 
of existing literature, suggesting exponential decay behavior 
$\langle\theta^2\rangle \sim e^{-\gamma t}$ in the large-time regime, 
where $\gamma>0$ is independent of $\kappa$. Finally, the analysis is 
extended to include a scalar source. For this case, the scalar variance
is found to become unbounded in forced-dissipative equilibrium in the 
asymptotic limit. 

The notion of diffusion anomaly involves subtle issues, which might be 
worthy of some clarification. As the product 
$\kappa\langle|\nabla\theta|^2\rangle$ would trivially vanish in the 
limit $\kappa\to0$ if $\langle|\nabla\theta|^2\rangle<\infty$, both the 
limits $\kappa\to0$ and $\langle|\nabla\theta|^2\rangle\to\infty$ are
key features. Furthermore, since $\langle|\nabla\theta|^2\rangle$ 
may not diverge in finite times, the limit $t\to\infty$ may also be
involved. As it turns out in the present study, 
$\langle|\nabla\theta|^2\rangle$ does not blow up in finite times. 
Hence, the limit $t\to\infty$ is indeed another key element of 
the problem. Now, the question is in what manner the limit $\kappa\to0$ 
should be considered to account for the associated divergence of 
$\langle|\nabla\theta|^2\rangle$ in infinite time.

This study takes an appropriate approach that answers the above 
question satisfactorily. Given fixed $\kappa>0$, let us consider the 
global maximum of the mean-square scalar gradients 
$\langle|\nabla\theta_T|^2\rangle$, where $\theta_T=\theta(\x,T)$ and 
$T$ denotes the corresponding time upon which this maximum is 
achieved. In general, $T$ depends on $\kappa$ and is expected to diverge 
in the limit $\kappa\to0$ as discussed in the preceding paragraph. The
present study does not attempt to determine how $T$ depends on $\kappa$,
except that the hypothetical scaling $T\approx\ln P_r$ is examined later 
on in \S\,3.4, where a related issue is discussed. The behavior of 
$\kappa\langle|\nabla\theta_T|^2\rangle$ 
is then monitored against all admissible distribution $\theta_T$ and its 
upper bounds are deduced in a $\kappa$-independent manner. Being derived 
in this way, these bounds are valid for all $\kappa$, including the limit 
$\kappa\to0$. Evidently, both limits $t\to\infty$ and
$\langle|\nabla\theta|^2\rangle\to\infty$ are being nested within the
limit $\kappa\to0$. Hence, the subsequent constraints on diffusion anomaly 
deduced from these bounds appropriately take into account the divergence 
of $\langle|\nabla\theta|^2\rangle$ in infinite time. In essence, this 
approach is consistent with the double limit 
$\lim_{\kappa\to0}\lim_{t\to\infty}$ taken in a selective way, in
the sense that a dynamical property (here being the peak mean-square 
gradients) is registered during the first limit $t\to\infty$ and then
monitored during the second limit $\kappa\to0$. Evidently, without such 
a selection, this double limit would yield a trivial result because no 
smooth scalar distribution would survive the limit $t\to\infty$ before 
$\kappa$ tends to zero. In passing, it is worth mentioning that the 
asymptotic regime of small but positive $\kappa$ and large but finite 
time, presumably $t\gg T$, is relevant in the study of exponential 
decay behavior mentioned in the opening paragraph of this section.

\subsection{Production and dissipation of scalar gradients}

The evolution of $\nabla\theta$ is governed by
\begin{eqnarray}
\label{gradients}
\partial_t\nabla\theta + (\u\cdot\nabla)\nabla\theta &=& 
\bomega\times\nabla\theta - (\nabla\theta\cdot\nabla)\u
+ \kappa\Delta\nabla\theta,
~~~~\nabla\cdot\u=0,
\end{eqnarray}
where $\bomega=\nabla\times\u$ is the vorticity. Apart from the ``Coriolis'' 
term $\bomega\times\nabla\theta$, whose sole effect is to rotate 
$\nabla\theta$ without changing its magnitude, Eq. (\ref{gradients}) 
resembles the three-dimensional vorticity equation in several aspects. 
For example, the gradient $\nabla\theta$ is advected by the velocity 
field $\u$ just as the vorticity $\bomega$. As another example, the term 
$(\nabla\theta\cdot\nabla)\u$, which is responsible for the production 
of scalar gradients, is an analogue of the vortex stretching term. 
Despite these apparent similarities, there are fundamental differences 
between ``genuine'' turbulence and passive scalar advection. First, 
the passive vector $\nabla\theta$ is irrotational whereas $\bomega$ is 
solenoidal. Second, $\theta$ is materially conserved while there exists
no such analogue in the vorticity dynamics. Third, genuine turbulence 
presumably becomes ``more turbulent'' in the inviscid limit while 
the limit of zero diffusivity in scalar advection does not make the 
advecting flows more effective stirrers. Among these differences, the 
third one appears to be most prominent, directly related to the 
active (and nonlinear) and passive (and linear) dynamical nature of 
$\bomega$ and $\nabla\theta$, respectively.  

Multiplying (\ref{gradients}) by $\nabla\theta$ and taking the domain average
of the resulting equation yields
\begin{eqnarray}
\label{bound1}
\frac{1}{2}\frac{d}{dt}\langle|\nabla\theta|^2\rangle  
&=& 
- \langle\nabla\theta\cdot(\nabla\theta\cdot\nabla)\u\rangle
- \kappa\langle|\Delta\theta|^2\rangle \nonumber\\
&\le&
\norm{\nabla\u}_\infty\langle|\nabla\theta|^2\rangle 
- \kappa\langle|\Delta\theta|^2\rangle,
\end{eqnarray}
where the advection and rotation terms identically vanish and the 
inequality is straightforward. Here $\norm{\nabla\u}_\infty$ 
denotes the supremum of $\nabla\u$ and, in this study, is assumed to
be bounded. This is the maximal degree of smoothness requiring of $\u$.
Given an initially smooth $\theta$, the manner in which 
$\langle|\nabla\theta|^2\rangle$ grows as $\kappa$ is decreased 
determines the outcome of the decay rate 
$2\kappa\langle|\nabla\theta|^2\rangle$. This depends primarily 
on how rapidly $\langle|\Delta\theta|^2\rangle$ grows. Note that even 
though $\langle|\Delta\theta|^2\rangle$ cannot be controlled a priori,
it does good rather than harm, effectively giving rise to the
boundedness of $\langle|\nabla\theta|^2\rangle$ for $\kappa>0$. This 
will become apparent in the next subsection.

A well-known fact which will be exploited later is that for sufficiently
small $\kappa$, $\langle|\nabla\theta|^2\rangle$ grows exponentially in 
time for an extended period in the initial stage until diffusive effects
become significant. For bounded $\norm{\nabla\u}_\infty$, it turns out 
that more rapid growth is not possible, even in the limit $\kappa\to0$. 
This is apparent from the inherent linearity of (\ref{gradients}) and 
can be stated more quantitatively in the following estimate, being 
obtained by integrating (\ref{bound1}), 
\begin{eqnarray}
\label{exp}
\langle|\nabla\theta|^2\rangle \le \langle|\nabla\theta_0|^2\rangle
\exp\{2\int_0^t\norm{\nabla\u}_\infty\,ds\} \le
 \langle|\nabla\theta_0|^2\rangle\exp\{2\Omega t\}.
\end{eqnarray}
Here $\theta_0=\theta(\x,0)$ and $\Omega$ is an upper bound for
$\norm{\nabla\u}_\infty$. Eq. (\ref{exp}) allows 
$\langle|\nabla\theta|^2\rangle$ to grow exponentially in time 
at a rate no greater than $2\Omega$ and forbids any possibility of 
a finite-time divergence. 

\subsection{Bounds for the variance decay rate}

For a preliminary estimate of the decay rate, let us consider (\ref{bound1}) 
in the form
\begin{eqnarray}
\label{anabound}
\frac{1}{2}\frac{d}{dt}\langle|\nabla\theta|^2\rangle  
&\le&
\norm{\nabla\u}_\infty\langle|\nabla\theta|^2\rangle 
- \kappa\langle|\Delta\theta|^2\rangle \nonumber\\
&=&
\frac{\langle|\Delta\theta|^2\rangle}{\langle|\nabla\theta|^2\rangle}
\left(\Omega\frac{\langle|\nabla\theta|^2\rangle^2}  
{\langle|\Delta\theta|^2\rangle}
- \kappa\langle|\nabla\theta|^2\rangle\right) \nonumber\\
&\le&
\frac{\langle|\Delta\theta|^2\rangle}{\langle|\nabla\theta|^2\rangle}
\left(\Omega\langle\theta^2\rangle 
- \kappa\langle|\nabla\theta|^2\rangle\right),
\end{eqnarray}
where (\ref{holder2}) with $\alpha=0$ has been applied to the second
equation. The term $\Omega\langle\theta^2\rangle$ is bounded and 
decreases monotonically. The relevant initial data is such that
$\langle|\nabla\theta|^2\rangle$ grows initially. This condition will 
be assumed throughout this section. Recall that $t=T$ denotes 
the time when $\langle|\nabla\theta|^2\rangle$ achieves its global 
maximum. This means that at $t=T$, $d\langle|\nabla\theta|^2\rangle/dt=0$. 
Eq. (\ref{anabound}) then implies that
\begin{eqnarray}
\label{anabound1}
\kappa\langle|\nabla\theta_T|^2\rangle&\le&\Omega\langle\theta_T^2\rangle.
\end{eqnarray}
This provides an upper bound for (half) the peak decay rate. Now, since 
$\langle|\nabla\theta|^2\rangle\le\langle|\nabla\theta_T|^2\rangle$ for 
all $t\ge0$ and $\langle\theta_T^2\rangle\le\langle\theta_0^2\rangle$, 
one can deduce that $\kappa\langle|\nabla\theta|^2\rangle\le\Omega
\langle\theta_T^2\rangle\le\Omega\langle\theta_0^2\rangle$ for all $t\ge0$.

Subject to certain conditions on the spectral distribution of $\theta_T$, 
which will be stated in due course, refined estimates for 
$\kappa\langle|\nabla\theta_T|^2\rangle$ can be derived. These conditions 
are admissible but not necessarily realizable, except possibly for the 
Batchelor spectrum, which has been supported by ample numerical and 
theoretical evidence. Hence, the results are {\it conditional} and as 
such should be interpreted accordingly. Applying (\ref{holder2}) to 
the second equation of (\ref{anabound}) yields
\begin{eqnarray}
\label{bound2}
\frac{1}{2}\frac{d}{dt}\langle|\nabla\theta|^2\rangle  
&\le&
\frac{\langle|\Delta\theta|^2\rangle}{\langle|\nabla\theta|^2\rangle}
\left(\Omega\frac{\langle|(-\Delta)^{\alpha/2}\theta|^2\rangle^{1/(1-\alpha)}}
{\langle|\nabla\theta|^2\rangle^{\alpha/(1-\alpha)}} 
- \kappa\langle|\nabla\theta|^2\rangle\right)\\
&=&
\frac{\langle|\Delta\theta|^2\rangle}{\langle|\nabla\theta|^2\rangle^
{1/(1-\alpha)}}\left(\Omega\langle|(-\Delta)^{\alpha/2}\theta|^2\rangle^
{1/(1-\alpha)} 
- \kappa\langle|\nabla\theta|^2\rangle^{1/(1-\alpha)}\right).\nonumber
\end{eqnarray}
In (\ref{bound2}), the terms 
$\Omega\langle|(-\Delta)^{\alpha/2}\theta|^2\rangle^{1/(1-\alpha)}$
and $\kappa\langle|\nabla\theta|^2\rangle^{1/(1-\alpha)}$ effectively
represent the relative strength of advective production and diffusive 
suppression of $\langle|\nabla\theta|^2\rangle$, respectively. Their 
balance shapes up the picture of variance decay. As will be seen 
presently, this balance is shifted in favor of the latter for steeper 
$\Theta(k)$. Applying the arguments in the derivation of (\ref{anabound1}) 
to (\ref{bound2}) yields
\begin{eqnarray}
\label{anabound2}
\kappa\langle|\nabla\theta_T|^2\rangle^{1/(1-\alpha)} \le
\Omega\langle|(-\Delta)^{\alpha/2}\theta_T|^2\rangle^{1/(1-\alpha)}.
\end{eqnarray}
Now, let us suppose that there exists an $\alpha_0\in(0,1)$ independent 
of $\kappa$ such that
\begin{eqnarray}
\label{condition} 
\langle|(-\Delta)^{\alpha/2}\theta_T|^2\rangle<\infty,
~~\mbox{for}~~\alpha\le\alpha_0.
\end{eqnarray}
It can be seen that if in the limit $\kappa\to0$, $\Theta_T(k)=\Theta(k,T)$ 
remains steeper than $k^{-1}$ for large $k$, such an $\alpha_0$ would exist. 
Given (\ref{condition}) and upon replacing $\alpha$ by $\alpha_0$, 
Eq. (\ref{anabound2}) becomes
\begin{eqnarray}
\kappa\langle|\nabla\theta_T|^2\rangle^{1/(1-\alpha_0)} \le
\Omega\langle|(-\Delta)^{\alpha_0/2}\theta_T|^2\rangle^{1/(1-\alpha_0)},
\end{eqnarray}
or equivalently
\begin{eqnarray}
\label{fbound}
\kappa\langle|\nabla\theta_T|^2\rangle \le
\Omega^{(1-\alpha_0)}\langle|(-\Delta)^{\alpha_0/2}\theta_T|^2\rangle
\kappa^{\alpha_0}.
\end{eqnarray}
This gives an upper bound for $\kappa\langle|\nabla\theta_T|^2\rangle$ 
in terms of $\kappa$, among other things, particularly $\alpha_0$. 
The divergence of $\langle|\nabla\theta_T|^2\rangle$ obeys 
the constraint
\begin{eqnarray}
\langle|\nabla\theta_T|^2\rangle \le
\Omega^{(1-\alpha_0)}\langle|(-\Delta)^{\alpha_0/2}\theta_T|^2\rangle
\kappa^{(\alpha_0-1)}.
\end{eqnarray}

A remarkable but not surprising feature of (\ref{fbound}) is the explicit 
dependence on $\alpha_0$, which could be termed the {\it degree of smoothness} 
of $\theta_T$. As it stands, Eq. (\ref{fbound}) reflects the obvious physical
connection between smoothness and diffusion that smoother scalar fields
diffuse more slowly. Note that (\ref{fbound}) also applies to 
the border case $\alpha_0=1$. This case is trivial, simply implying that 
$\kappa\langle|\nabla\theta_T|^2\rangle$ scales as $\kappa$ if
$\langle|\nabla\theta_T|^2\rangle<\infty$. Note also that the conditional 
bound (\ref{fbound}) is an improvement over the result of Tran and 
Dritschel$^{23}$ for passive scalar in two dimensions and active 
potential vorticity in various geophysical fluid models. A similar
improvement is also possible for their case of enstrophy dissipation 
in two-dimensional Navier--Stokes turbulence, for which the assumption 
$\norm{\nabla\u}_\infty<\infty$ (which is not justifiable) is replaced
at a minimal cost by the natural condition of finite vorticity (which is
materially conserved). By 
replacing $\theta$ in (\ref{holder2}) and (\ref{condition}) by the 
vorticity and applying the resulting equations to their Eq. (3.2), one 
can deduce an upper bound for the maximum enstrophy dissipation rate that 
scales as $\nu^{\alpha_0/(2-\alpha_0)}$. This scaling converges to zero 
more slowly than the scaling $\kappa^{\alpha_0}$. The slower convergence 
is the cost for relaxing the condition $\norm{\nabla\u}_\infty<\infty$.

On physical grounds, $\Theta_T(k)$ can become shallower as $\kappa$ is 
decreased. This means that condition (\ref{condition}) may not be fulfilled, 
i.e. there may exist no $\alpha>0$ independent of $\kappa$ such that 
$\langle|(-\Delta)^\alpha\theta_T|^2\rangle<\infty$. If this is the case, 
the scalar field may be said to become virtually singular. This includes 
the Batchelor spectrum $\Theta(k)\propto k^{-1}$, which has been 
believed to be the correct scaling of the so-called viscous-convective 
range, i.e. the range between the energy dissipation scale and the scalar 
diffusion scale. A plausible scenario is that $\Theta_T(k)$ becomes 
increasingly shallower, approaching the limiting scaling $k^{-1}$ in the 
limit $\kappa\to0$. For this case, applying (\ref{holderer}) to 
the second equation of (\ref{anabound}) yields
\begin{eqnarray}
\label{bound3}
\frac{1}{2}\frac{d}{dt}\langle|\nabla\theta|^2\rangle  
&\le&
\frac{\langle|\Delta\theta|^2\rangle}{\langle|\nabla\theta|^2\rangle}
\left(\frac{\Omega\langle\theta^2\rangle}{\ln(k_1/k_2)} 
- \kappa\langle|\nabla\theta|^2\rangle\right) \nonumber\\
&\le&
\frac{\langle|\Delta\theta|^2\rangle}{\langle|\nabla\theta|^2\rangle}
\left(\frac{\Omega\langle\theta^2\rangle}{\ln(k_\kappa/k_\nu)} 
- \kappa\langle|\nabla\theta|^2\rangle\right),
\end{eqnarray}
where $k_1$ and $k_2$ have been replaced by the viscous dissipation 
wave number $k_\nu$ and the diffusion wave number $k_\kappa$, respectively.
Following Tran and Dritschel,$^{23}$ let us identify the ratio 
$k_\kappa^2/k_\nu^2$ with the Prandtl or Schmidt number $P_r$. By 
substituting $P_r=k_\kappa^2/k_\nu^2$ into (\ref{bound3}), one obtains
\begin{eqnarray}
\frac{1}{2}\frac{d}{dt}\langle|\nabla\theta|^2\rangle  
&\le&
\frac{\langle|\Delta\theta|^2\rangle}{\langle|\nabla\theta|^2\rangle}
\left(\frac{2\Omega\langle\theta^2\rangle}{\ln P_r} 
- \kappa\langle|\nabla\theta|^2\rangle\right).
\end{eqnarray} 
It follows that 
\begin{eqnarray}
\label{gbound}
\kappa\langle|\nabla\theta_T|^2\rangle &\le& 
\frac{2\Omega\langle\theta_T^2\rangle}{\ln P_r}.
\end{eqnarray}
Eq. (\ref{gbound}) implies an extraordinarily weak dependence of 
$\kappa\langle|\nabla\theta_T|^2\rangle$ on $P_r$. Nevertheless, 
$\kappa\langle|\nabla\theta_T|^2\rangle$ vanishes in the limit 
$P_r\to\infty$. Similar lines of calculation$^{23}$ for the case of 
enstrophy decay in two-dimensional Navier--Stokes turbulence have 
found that the enstrophy dissipation rate scales with the Reynolds
number $Re$ as $(\ln Re)^{-1/2}$. Again, this relatively slower 
convergence to zero (as compared with $(\ln P_r)^{-1})$ is the cost
for relaxing the condition $\norm{\nabla\u}_\infty<\infty$.

\subsection{Discussion}

The results in the preceding subsection cover a broad class of scalar 
distribution corresponding to no anomalous diffusion. This class of 
regular distribution includes power-law spectra no shallower than 
$k^{-1}$. The type of ``singular'' distribution that would give rise 
to diffusion anomaly would feature a significant fraction of 
$\langle\theta^2\rangle$ ``cascading'' to the vicinity of an 
ever-increasing diffusion wave number $k_\kappa\propto\kappa^{-1/2}$ as 
$\kappa\to0$. The decay rate would then be proportional to this fraction. 
As can be seen from the estimates of \S\,2, this fraction would have to 
be greater than that allowed by the Batchelor $k^{-1}$ spectrum. This is 
consistent with recent results$^{1,7,14,19}$ suggesting spectra shallower 
than $k^{-1}$. Note that for such spectra, virtually all (not just a 
fraction) of $\langle\theta^2\rangle$ is at the small scales. 

A number of studies$^{8,13,17,20}$ have suggested that in the large-time
regime, the variance decays as 
$\langle\theta^2\rangle\propto e^{-\gamma t}$, where $\gamma>0$ is 
independent of $\kappa$. The exponential decay rate $\gamma$ differs 
from the usual decay rate $2\kappa\langle|\nabla\theta|^2\rangle$ and 
is given by
\begin{eqnarray}
\label{erate}
\gamma = - \frac{1}{\langle\theta^2\rangle}
\frac{d}{dt}\langle\theta^2\rangle
=\frac{2\kappa\langle|\nabla\theta|^2\rangle}{\langle\theta^2\rangle}.
\end{eqnarray}
In the presence of diffusion anomaly (the exceptional case of singular 
distribution mentioned above), this large-time regime presumably 
corresponds to $t\approx T(\kappa_0)$ for some small $\kappa=\kappa_0$, 
beyond which $\kappa\langle|\nabla\theta_T|^2\rangle$ becomes independent 
of $\kappa$. If $\gamma$ stays approximately constant for a period of time,
then the ratio $\langle|\nabla\theta|^2\rangle/\langle\theta^2\rangle$ 
is required to remain approximately unchanged, i.e. both 
$\langle|\nabla\theta|^2\rangle$ and $\langle\theta^2\rangle$ decay 
approximately at the same rate $\gamma$. In the absence of diffusion 
anomaly, $\gamma$ vanishes in the same manner as the usual decay rate 
for $t\approx T$. For example, for the Batchelor spectrum, Eqs. 
(\ref{gbound}) and (\ref{erate}) imply that the exponential decay rate 
at $t=T$, denoted by $\gamma_T$, satisfies
\begin{eqnarray}
\gamma_T \le \frac{4\Omega}{\ln P_r}.
\end{eqnarray}
Although $\gamma_T$ depends on $P_r$ and vanishes in the limit 
$P_r\to\infty$, in principle $\gamma$ might become sizable for 
$t\gg T$, provided that the ratio 
$\langle|\nabla\theta|^2\rangle/\langle\theta^2\rangle$
has increased by a $\ln P_r$-fold since $t=T$. This requires 
of the variance itself a decrease of at least a $\ln P_r$-fold, i.e. 
$\langle\theta^2\rangle\le\langle\theta_T^2\rangle/\ln P_r$, because 
$\langle|\nabla\theta|^2\rangle$ also decays (at least on average if not 
monotonically). Note that an increase of 
$\langle|\nabla\theta|^2\rangle/\langle\theta^2\rangle$ requires that 
$\langle\theta^2\rangle$ decay relatively more rapidly than 
$\langle|\nabla\theta|^2\rangle$. This means that the advective 
production of scalar gradients remains considerably active throughout
the regime $t>T$. 

An interesting problem arises from this analysis is that regardless of
the realizability of diffusion anomaly, the cumulative decay, up to 
$t=T$, defined by
\begin{eqnarray}
\label{cumulative}
\langle\theta_0^2\rangle - \langle\theta_T^2\rangle = 
2\kappa\int_0^{T}\langle|\nabla\theta|^2\rangle\,dt
\end{eqnarray}
may remain non-zero and be independent of $\kappa$. A similar problem 
on the cumulative decay of enstrophy in freely evolving two-dimensional 
turbulence has been briefly addressed in a recent numerical study,$^{26}$ 
illustrating the absence of enstrophy dissipation anomaly. There, it was 
observed that by the 
time $T(Re)$ of peak enstrophy dissipation, the enstrophy loss was 
significant and, more importantly, roughly the same for a range of 
Reynolds numbers $Re$. It was also noted that $T(Re)$ increased with 
$Re$, probably as $\ln Re$. In the light of these results, it seems 
plausible that the cumulative decay, defined by (\ref{cumulative}), 
remains non-zero in the limit $\kappa\to0$. For a quantitative sense 
of how this might be the case, let us consider the following picture, 
which is in accord with the above observations for the decay of 
two-dimensional enstrophy. In accord with (\ref{exp}), 
$\langle|\nabla\theta|^2\rangle$ is expected to grow exponentially
for a good part of the time interval $[0,T]$, before diffusive 
suppression becomes significant toward the end. Hence, one may write
$\langle|\nabla\theta|^2\rangle\approx\langle|\nabla\theta_0|^2\rangle
\exp\{c_1\Omega t\}$, where $c_1>0$ is a dimensionless constant. Given 
this growth, the requirement of $T$ for the right-hand side of 
(\ref{cumulative}) to be independent of $\kappa$ is 
$T\approx c_1^{-1}\Omega^{-1}\ln(c_2P_r)$, where $c_2>0$ is another
dimensionless constant. Indeed, by substituting these into 
(\ref{cumulative}), the integral on the right-hand side can be readily 
evaluated, and the result is $\propto P_r$. This effectively cancels 
out the diffusion coefficient in (\ref{cumulative}). Thus the cumulative 
decay could be non-zero and independent of $\kappa$ in the limit 
$\kappa\to0$. In the absence of diffusion anomaly, one may expect this 
picture to be most plausible for the Batchelor spectrum. The reason is 
that diffusive suppression of scalar gradients becomes relatively 
stronger (as compared with their advective production) for steeper 
spectra, as suggested by the analysis in \S\,3.2. This means that for 
power-law spectra no shallower than $k^{-1}$, the $k^{-1}$ spectrum is 
most favorable for the gradient production and therefore highly 
likely to return the most cumulative decay.

\subsection{The forced case}

This subsection considers the dynamical picture of forced-dissipative 
equilibrium in the limit $\kappa\to0$, where the scalar variance is 
perpetually replenished at the large scales by a scalar source. In the 
presence of such a source, the absence of diffusion anomaly would be 
catastrophic, in the sense that the scalar field would diverge, 
particularly at the injection scales. It turns out that for power-law 
spectra, $\langle\theta^2\rangle$ diverges, regardless of whether or 
not diffusion anomaly is realizable. The classical spectrum 
$\Theta(k)\propto k^{-1}$ is the steepest one that could remain bounded 
while still satisfying the condition $\langle\theta^2\rangle\to\infty$. 
Steeper spectra would necessarily grow without bound, i.e. blowup at 
large scales.

With the addition of a scalar source $f$ to the right-hand side of the 
advection-diffusion equation in (\ref{governing}), Eqs. (\ref{decay}) 
and (\ref{bound1}) become
\begin{eqnarray}
\label{decay1}
\frac{1}{2}\frac{d}{dt}\langle\theta^2\rangle &=& 
-\kappa\langle|\nabla\theta|^2\rangle + \langle\theta f\rangle
\end{eqnarray}
and
\begin{eqnarray}
\label{bound4}
\frac{1}{2}\frac{d}{dt}\langle|\nabla\theta|^2\rangle  
&\le&
\frac{\langle|\Delta\theta|^2\rangle}{\langle|\nabla\theta|^2\rangle}
\left(\Omega\frac{\langle|\nabla\theta|^2\rangle^2}
{\langle|\Delta\theta|^2\rangle}
+\frac{\langle\nabla \theta\cdot\nabla f\rangle\langle|\nabla\theta|^2\rangle}
{\langle|\Delta\theta|^2\rangle} 
- \kappa\langle|\nabla\theta|^2\rangle\right),
\end{eqnarray}
respectively. Let us consider a persistent source as described above, i.e.
$\langle\theta f\rangle>0$, and forced-dissipative equilibrium dynamics,
i.e. $\kappa\langle|\nabla\theta|^2\rangle=\langle\theta f\rangle$.
These can be understood in an average sense and are assumed for all $P_r$. 
Suppose that in the limit $\kappa\to0$, $\langle\theta^2\rangle$ remains 
bounded. Then for power-law scaling, $\Theta(k)$ is steeper than $k^{-1}$. 
From the above calculation for the unforced case, the first term in 
the brackets on the right-hand side of (\ref{bound4}) vanishes as 
$\langle|\nabla\theta|^2\rangle\to\infty$. The second term also vanishes 
for a broad class of scalar sources $f$, particularly those satisfying 
$\langle|\nabla f|^2\rangle<\infty$. It follows that for 
$\langle\theta f\rangle>0$, the assumed forced-dissipative equilibrium 
$\kappa\langle|\nabla\theta|^2\rangle=\langle\theta f\rangle$
cannot be achieved. This means that such an equilibrium requires
$\langle\theta^2\rangle\to\infty$. The spectrum $\Theta(k)=ck^{-1}$, 
extending to infinity and having non-diminishing $c$, can be seen as a 
possible solution, corresponding to a logarithmic divergence of the 
scalar variance toward the small scales. This singular distribution is 
special and anomalously diffused. Other cases of blowup via unbounded 
spectra steeper than $k^{-1}$ are also permitted. These may or may not 
be accompanied by diffusion anomaly. Note that non-power-law spectra
having significant variance at ever-smaller scales and 
$\langle\theta^2\rangle<\infty$ cannot be ruled out. Similar arguments 
have been used by Tran$^{27}$ for the problem of enstrophy dissipation 
in forced-dissipative two-dimensional turbulence (in an essentially 
unbounded domain). There, the vorticity was found to diverge in the 
inviscid limit.

\section{Conclusion}

In conclusion, this study has derived upper bounds for the peak decay rate 
of the variance of a passive scalar $\theta$ advected by three-dimensional 
flows having bounded velocity gradients $|\nabla\u|<\infty$. These bounds 
are expressible in terms of controlled physical parameters and conditional 
on the distribution of the scalar field $\theta_T=\theta(\x,T)$ at the time 
of peak decay $t=T$. It has been found that in the limit of vanishingly 
small diffusivity $\kappa$, the peak decay rate vanishes as 
$\kappa^{\alpha_0}$ if there exists an $\alpha_0\in(0,1]$ independent of 
$\kappa$ such that $\langle|(-\Delta)^{\alpha/2}\theta_T|^2\rangle<\infty$ 
for $\alpha\le\alpha_0$. This condition is satisfied if at $t=T$, the 
variance spectrum $\Theta_T(k)$ is steeper than $k^{-1}$ at the spectral 
tail, i.e. at large wave numbers $k$. Such a positive $\alpha_0$ is not 
known to exist as numerical simulations$^{28}$ indicate an increasing 
roughness of the scalar field with increasing $P_r$. If this hypothesis 
turns out to be false, a plausible scenario consistent with Batchelor's 
theory is that $\Theta(k)$ becomes increasingly shallower for smaller 
$\kappa$, approaching the Batchelor scaling $k^{-1}$ in the limit 
$\kappa\to0$. For this classical spectrum, the peak decay rate also 
vanishes, albeit more slowly --- like $(\ln P_r)^{-1}$. These results rule 
out anomalous diffusion for a broad range of $\Theta_T(k)$, including 
power-law spectra no shallower than $k^{-1}$. The implication is that a 
non-vanishing decay rate in the limit $\kappa\to0$ necessarily requires 
the variance at small scales to be greater than that allowed by the 
Batchelor spectrum. The absence of anomalous diffusion in the above cases 
can be seen as a consequence of strong diffusive suppression of scalar 
gradients, which are relatively weakly produced by advection. In the 
presence of a scalar source, $\langle\theta^2\rangle$ has been found to 
diverge in the limit $\kappa\to0$ if $\Theta(k)$ obeys power-law scalings. 
Hence, spectra steeper than $k^{-1}$ would necessarily grow without bound. 
The classical $k^{-1}$ spectrum, being bounded and accompanied by diffusion 
anomaly, is a possible solution, corresponding to a logarithmic divergence 
of the variance toward the small scales. Interestingly, a diverging scalar 
field would be necessary for transferring the injected scalar variance 
to the small scales for disposal. 

The present results have been derived on the assumption of bounded 
velocity gradients. This assumption is plausible for the Batchelor 
turbulence regime, where the Prandtl number $P_r$ tends to infinity 
while the Reynolds number $Re$ remains bounded. In fact, there exists 
neither theoretical nor numerical evidence suggesting otherwise. The 
case where $Re\to\infty$ and $P_r>0$ is interesting and presumably 
much more involved. In this case, one would expect the advecting flows 
to become increasingly ``rougher'', possibly leading to unbounded 
velocity gradients in the limit $Re\to\infty$. Intuitively, diffusion 
anomaly is more plausible for this case because the scalar gradient 
production term can no longer be controlled a priori. This is no 
surprise as the original notion of energy dissipation anomaly is 
associated with the limit of infinite Reynolds number.

\vspace{.5cm}

\noindent$^1$T. M. Antonsen, Jr. Zhencan Fan, E. Ott, and E. Garcia--Lopez,
``The role of chaotic orbits in the determination of power spectra of passive 
scalars,'' Phys. Fluids {\bf 8}, 3094 (1996). 

\noindent$^2$E. Balkovsky and A. Fouxon, ``Universal long-time properties of
Lagrangian statistics in the Batchelor regime and their application to the
passive scalar problem,'' Phys. Rev. E {\bf 60}, 4164 (1999).

\noindent$^3$G. K. Batchelor, ``Small-scale variation of convected 
quantities like temperature in turbulent fluid. Part 1. General discussion and
the case of small conductivity,'' J. Fluid Mech. {\bf 5}, 113 (1959).

\noindent$^4$G. K. Batchelor I. A. Howell, and A. A. Townsend, ``Small-scale 
variation of convected quantities like temperature in turbulent fluid. Part 2. 
The case of large conductivity,'' J. Fluid Mech. {\bf 5}, 134 (1959).

\noindent$^5$M. Chertkov, G. Falkovich, and I. Kolokolov, ``Intermittent
dissipation of a passive scalar in turbulence,'' Phys. Rev. Lett. {\bf 80},
2121 (1998).

\noindent$^6$S. Corrsin, ``On the spectrum of isotropic temperature fluctuation
in isotropic turbulence,'' J. Appl. Phys. {\bf 22}, 469 (1951).

\noindent$^7$D. R. Fereday and P. H. Haynes,
``Scalar decay in two-dimensional chaotic advection and Batchelor-regime
turbulence,'' Phys. Fluids {\bf 16}, 4359 (2004). 

\noindent$^8$P. H. Haynes and J. Vanneste, ``What controls the decay rate
of passive scalar in smooth random flows?'' Phys. Fluids {\bf 17}, 097103 
(2005).

\noindent$^9$R. H. Kraichnan, ``Small-scale structure of a scalar field
convected by turbulence,'' Phys. Fluids {\bf 11}, 945 (1968).

\noindent$^{10}$W. Liu, ``Does a fast mixer really exist?'' Phys. Rev. E
{\bf 72}, 016312 (2005). 

\noindent$^{11}$A. D. Majda, ``The random uniform shear layer: an explicit
example of turbulent diffusion with broad tail probability distributions,'' 
Phys. Fluids A {\bf 5}, 1963 (1993). 

\noindent$^{12}$A. M. Obukhov, ``The structure of the temperature field in 
a turbulent flow,'' Izv. Akad. Nauk. SSSR, Ser. Geophys. {\bf 13}, 58 (1949).

\noindent$^{13}$R. T. Pierrehumbert, ``Tracer microstructure in the large-eddy
dominated regime,'' Chaos, Solitons Fractals. {\bf 4}, 1091 (1994).

\noindent$^{14}$A. A. Schekochihin, P. H. Haynes, and S. C. Cowley,
``Diffusion of passive scalar in a finite random flow,'' Phys. Rev. E {\bf 70},
046304 (2004).

\noindent$^{15}$K. R. Sreenivasan, ``The passive scalar spectrum and the 
Obukhov--Corrsin constant,'' Phys. Fluids {\bf 8}, 189 (1996).

\noindent$^{16}$K. R. Sreenivasan, ``On local isotropy of passive scalars in 
turbulent shear flows,'' Proc. Roy. Soc. A {\bf 434}, 165 (1991).

\noindent$^{17}$J. Sukhatme and R. T. Pierrehumbert, ``Decay of passive scalars
under the action of single scale smooth velocity fields in bounded 
two-dimensional domains: from non-self-similar probability distribution 
function to self-similar Eigenmodes,'' Phys. Rev. E {\bf 66}, 056302 (2002).

\noindent$^{18}$C. V. Tran, ``An upper bound for passive scalar diffusion in
shear flows,'' Phys. Fluids {\bf 19}, 068104 (2007).

\noindent$^{19}$Y. K. Tsang, T. M. Antonsen, and E. Ott, ``Exponential decay
of chaotically advected passive scalars in the zero diffusivity limit,''
Phys. Rev. E {\bf 71} 066301 (2005).

\noindent$^{20}$A. Wonhas and J. C. Vassilicos, ``Mixing in fully chaotic
flows,'' Phys. Rev. E {\bf 66}, 051205 (2002).

\noindent$^{21}$P. K. Yeung, Shuyi Xu, and K. R. Sreenivasan, ``Schmidt number
effects on turbulent transport with uniform mean scalar gradient,''
Phys. Fluids {\bf 14}, 4178 (2002).

\noindent$^{22}$P. A. Davidson, ``Turbulence: An Introduction for Scientists 
and Engineers,'' Oxford University Press (2004).

\noindent$^{23}$C. V. Tran and D. G. Dritschel, ``Vanishing enstrophy 
dissipation in two-dimensional Navier--Stokes turbulence in the inviscid 
limit,'' J. Fluid Mech. {\bf 559}, 107 (2006).

\noindent$^{24}$C. V. Tran, ``Nonlinear transfer and spectral distribution of 
energy in $\alpha$ turbulence,'' Physica D {\bf 191}, 137 (2004).

\noindent$^{25}$C. V. Tran and T. G. Shepherd, ``Constraints on the spectral 
distribution of energy and enstrophy dissipation in forced two-dimensional 
turbulence,'' Physica D {\bf 165}, 199 (2002). 

\noindent$^{26}$D. G. Dritschel, C. V. Tran and R. K. Scott, ``Revisiting
Batchelor's theory of two-dimensional turbulence,'' J. Fluid Mech. {\bf 591},
379 (2007).

\noindent$^{27}$C. V. Tran, ``Constraints on inertial range scaling laws in
forced two-dimensional Navier--Stokes turbulence,'' Phys. Fluids {\bf 19}, 
108109 (2007).

\noindent$^{28}$J. Schumacher and K. R. Sreenivasan, ``Geometric features 
of the mixing of passive scalars at high Schmidt numbers,'' Phys. Rev. Lett. 
{\bf 91}, 174501 (2003).

\end{document}